\documentclass[conference]{IEEEtran}
\IEEEoverridecommandlockouts
\usepackage{cite}
\usepackage{amsmath,amssymb,amsfonts}
\usepackage{algorithmic}
\usepackage{graphicx}
\usepackage{textcomp}
\usepackage{xcolor}
\usepackage{verbatim}
\def\BibTeX{{\rm B\kern-.05em{\sc i\kern-.025em b}\kern-.08em
    T\kern-.1667em\lower.7ex\hbox{E}\kern-.125emX}}

\begin{document}

\title{Smart Mat Used for Prevention of Hospital-Acquired Pressure Injuries\\
}


\author{Lanhao Gao, Zihuai Lin\\
	School of Electrical and Information Engineering, The University of 
	Sydney, Australia\\
	Emails:	zihuai.lin@sydney.edu.au. 
}

\maketitle

\begin{abstract}
This work develops a smart mat for monitoring body positions. We use Velostat as a force sensor resistance (FSR) to construct a sensor matrix over the mat to receive the pressure distribution of the patient's body, and then upload the processed distribution information to the PC for data visualization through Arduino. Data visualization on the PC side is compiled through Python language to realize the functions of patient body pressure distribution monitoring, long-term pressure alarm and posture prediction. The purpose of this work is to relieve the work stress on medical staff caused by pressure injuries during the treatment and care of patients during the pandemic. This paper includes the literature review on similar previous works and combines the test results to design the structure and circuit of the smart mat. 
\end{abstract}
\smallskip
\begin{IEEEkeywords}
smart mat, velostat, FSR, pressure injuries
\end{IEEEkeywords}

\section{Introduction}\label{sec:introduction}

The use of the Internet of Things is growing steadily over the years. It is expected that by 2025, there will be approximately 27 billion connected IoT devices \cite{IoTanalysis}. At present, the Internet of Things is one of the main promoters of technological innovation and one of the areas with greater potential for social and economic transformation  \cite{leng2020implementation,IoT_FD,RF_energy1}. 

Through a network of sensors and actuators connected to a wireless network \cite{NC1,NC2,NC3,WRN,distributedRaptor,JNCC,RCRC,NC_book}, the operator has the power to remotely gather data. Alternatively, actuators could be programmed to actuate automatically according to values reported by the sensor.

In recent years, various smart mats with many kinds of functions for different fields have appeared on the market, and the smart mats for people’s livelihood are mainly used to improve people’s sleep quality and detect the health status of humans’ bodies. 
The data collected by the mat's sensors will be uploaded to the user interface, and the mat's data display and function control are completely reliant on the Graphical user interface.
Although these mats can detect human sleep data so that people can improve sleep quality based on those data, the relevant data cannot accurately meet medical demands, especially in preventing hospital-acquired pressure injuries. The prevention of hospital-acquired pressure injuries has been one of the most difficult aspects of hospital treatment for a very long time, and the challenges in prevention and management have placed a significant burden on medical staff.
The occurrence of pressure injuries not only brings great pain to patients and their families but also increases the cost of nursing care. According to statistics from public hospitals in Australia, 121,645 cases of hospital-acquired pressure injuries resulted in about 982 million AU dollars in additional medical expenses and 524,661 days of bed loss between 2012 and 2013.

This work is dedicated to developing an intelligent mat that could provide high accuracy to meet the requirement of preventing pressure injuries. In the process of hospital care, it is difficult for medical staff to find the specific position of pressure injuries, and regular inspections make the care process cumbersome. Thus, the primary goal of the mat designed in this work is to assist nursing staff by visualising the occurrence position of pressure injury in the patient body.  

The smart mat designed in this paper focus on relieving the pressure on nursing work caused by pressure sores during the pandemic through the visualization of the pressure of the patient's body and reminding the medical staff to help the patient move at a particular time when the body is under pressure for a long time. In addition, for critically ill patients who are in isolation, medical staff need to obtain as much information about the patient’s physical signs as possible and remotely receive alerts about pressure sores that may occur to further help patients in treatment and prevention.

\section{Background\label{cha:litreivew}}

\subsection{Pressure injury}

The pressure injury, as a long-term and painful pressure-induced illness, is one possible result of long-term hospitalisation, and this type of illness usually occurs in patients who are unable to move or lose consciousness due to other illnesses. In different regions, these kinds of injuries are called bedsores, pressure ulcers, etc. 

\subsubsection{Four stage classification of pressure injury}

The pressure injury is usually one type of local damage to the skin and subcutaneous soft tissue caused by medical or other equipment in places of bone protrusion, the manifestation of this injury is intact skin or open ulcers and may be accompanied by pain. The National Pressure Ulcer Advisory Panel (NPUAP) was modified the 'pressure ulcer' to 'pressure injury' and updated its pressure classification in the statement which is published in 2016 \cite{24}. 

In the latest pressure injury stage classification system, pressure injury more accurately describes the pressure injury at the intact or ulcerated skin. There are many causes of stress injuries, such as stress, humidity and temperature in parts of the body, but the tolerance of human soft tissues to these forces may be changed by factors such as micro-environment, nutrition, comorbidities and soft tissue conditions \cite{24}.

The National Pressure Ulcer Advisory Committee has divided the occurrence of pressure ulcers into four stages in detail. When the patient is in the first stage, the skin surface will present normally as intact skin without traces, but when the finger presses on the local skin, once the red marks appear on the surface of skin and cannot recover quickly, the patient's skin is likely to have deep tissue pressure damage, and in the original state, the feeling, temperature, and skin firmness are abnormal when pressing at this time. However, in the second stage of pressure injury, part of the skin has been lost and part of the dermis is exposed where the patient’s skin is missing \cite{5}. 

When in the second stage, the patient’s wound is still active, which surface is always keeping moist, and the color of the epidermis is pink or red. Although serous blisters appear on the skin lesions, fat and deeper tissues are not seen at this stage. 

The third stage is called full-thickness skin loss, which means that the patient's skin is severely lacking and the fat also can be seen obviously in the wound, at the same time, the granulation tissue and the skin edge are rolled inward. 

Finally, when the patient's skin is excessively damaged and the full thickness of the skin and tissues fall off, it is called the fourth stage. The ulcers produced in the fourth stage are exposed to the air, which damages the fascia, muscles, ligaments, bones and other tissues so that they are in direct contact with the air, so carrion and eschar can be seen \cite{5}. The pressure injury in the fourth stage will be accompanied by a strong irritating stench, which requires medical staff not only need to care and treat the wound during the treatment process, but also the management of odor is also one of the most important tasks.

In addition, when the whole skin tissue of the patient is missing at a local location, the degree of tissue damage inside the wound cannot be distinguished due to the coverage of decayed flesh and eschar, which makes it impossible to determine the third or fourth stage. At this time, NPUAP calls it an indistinguishable stage. 

\subsubsection{Causes of occurrence}
    
The various stages of pressure injury are caused by certain external force factors that cause different degrees of damage to the skin under fixed conditions. First, pressure injury can be defined as tissue ischemia caused by applying external forces damage such as shear and pressure \cite{6}. 
    
This work takes hospital-acquired pressure injuries as the main prevention target and needs to analyze the main factors that cause this type of pressure ulcers. The cause of pressure injury is directly related to the three forces of pressure, shear and friction. First, when the pressure on the skin of the patient is uneven and deform the tissues change for a long time, for example, the supporting surface on the body is uneven, the part of the skin cannot survive due to vascular ischemia \cite{6}. Moreover, when the force moving in different directions is applied to the surface of the tissue, it will cause damage caused by shear force \cite{7}. Scenes that cause injury caused by shearing force, such as at the skin support surface, because the friction of the body hinders the movement of the body, the tissue between the skin and the bone is twisted, and any blood vessels and tissues passing through the area are sheared. Therefore, the probability of ulcers caused by shear is higher than the chance due to the pressure \cite{7}. Finally, friction as a necessary condition for causing shear damage is more likely to cause skin pressure ulcers, and it usually needs to be matched with others forces to accelerate the skin rupture \cite{6}. 
    
In addition to the various pressure factors that can directly cause pressure ulcers, there are actually many other equally important indirect factors that may also become one of the triggers of pressure ulcers or increase the risk in happening of pressure injury. First of all, patients with the inconvenience caused by certain diseases or in the recovery period may have a relevant high possibility of acquiring pressure ulcers without timely care \cite{7}. Another important indirect factor is that age-related physiological characteristics lower the threshold for patients to obtain pressure injury under the same conditions \cite{7}. As the increasing of the age, the skin of patients not only loses the amount of elasticity, but also the blood and subcutaneous fat inside the skin will be relatively reduced, which makes elderly patients more susceptible to pressure ulcers. Moreover, many patients have urinary incontinence or intestinal incontinence accompanied by other serious illnesses, and this incontinence will increase the humidity in certain areas of the skin, thereby increasing the risk of forming pressure ulcers \cite{6}.
    
In conclusion, the main factors that can contribute to the formation of pressure injuries are pressure, humidity, and the health status of the patient’s skin. At the same time, these factors have an inseparable relationship with time because they all cause more and more damage to the patient’s skin with the increase of time.

\subsubsection{Pressure injuries occurrence in pandemic}

During the pandemic of the past two years, many countries around the world are under tremendous medical pressure due to the appearance of viruses, and in the period of the treatment, the considerable stress of medical care makes it easy for people to forget the existence of pressure sores. In particular, older and critically ill patients who have mobility impairments and are in isolation may have a significantly increased probability of stress injuries due to a lack of timely and effective care.

Coronavirus can cause illnesses ranging from colds to critical vital signs. Acute Respiratory Distress Syndrome (ARDS) is one of the common illnesses, which may lead to respiratory failure and lung damage in patients. Therefore, most patients with severe coronavirus need to use a ventilator in the intensive care unit and through the prone position to assist breathing \cite{29}. When the patient is in the mechanical ventilation stage, the length of the prone position treatment usually exceeds 12 hours in one day, which makes the patient's face one of the areas with the highest risk of pressure injury \cite{29}.

    \vspace*{1\baselineskip}
    \begin{figure}[h]  

    \centering  

    \includegraphics[width=0.8\linewidth]{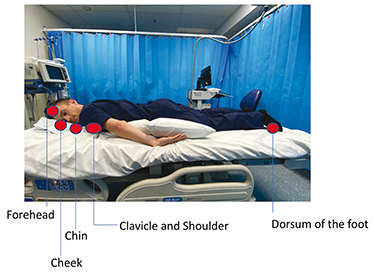} 

    \caption{Areas of potential pressure injury formation when lying in a prone position \cite{29}.}

    \end{figure}   

Generally, research on the prevention of pressure ulcers is aimed at the parts of the body in the supine position, but the emergence of the coronavirus has caused many patients to use the prone position and ventilator to make that the face become one of the high-risk areas. Therefore, the detection and prevention of facial pressure ulcers are also essential.
    
\subsubsection{Preventative measure}
    
The traditional treatment of pressure injury uses a variety of corresponding methods according to the different stages of the patient's wound. However, its fundamental and most effective method is to prevent it in the initial stage \cite{7}.
    
The most effective stage of the hospital's intensive care plan for the treatment of stress injuries is prevention. Patients who have been in a wheelchair for a long time or are bedridden due to severe illness are experiencing a high risk of acquiring pressure ulcers in their treatment period in the hospital \cite{7}. Moreover, post-care is a disaster level for medical staff and their families once these people get pressure injuries. Therefore, preventing pressure ulcers in hospital nursing work has become an important step. 
    
The prevention of pressure injuries is generally divided into three parts: assistance, inspection and nutrition. Because pressure injuries are mainly caused by long-term or instant pressure on a certain skin, it is necessary for nursing staff to assist patients in changing positions regularly and frequently \cite{9}. The change of body position and posture has always been one of the most popular methods to prevent pressure ulcers. Therefore, health service agencies will require medical care relevant staff to have various capabilities in risk monitoring, prevention plans and pressure injury management plans during labour training in order to improve their understanding of pressure injuries and proficiency in work operations. Generally, a wheelchair user needs to change position at least every half an hour. At the same time, a patient confined on the bed must be assisted to complete actions like move, turn over and other operations at least every two hours \cite{7}. In addition, the specific practices of pressure injury monitoring in the document of Australian Safety and Quality in Health Care \cite{4} require clinicians and nursing staff to develop a comprehensive care plan that provides pressure injury prevention strategies and regular detection and reassessment for patients' skin health status. Furthermore, patients need to supplement nutrition every day to enhance the health of the skin. The blood circulation in the skin and the maintenance of elasticity can reduce the occurrence of pressure sores as much as possible \cite{4}.
    
\subsubsection{Current products}
    
People at high risk of getting pressure ulcers, such as bedridden, elderly and obese people, not only need to adjust the lying and sitting posture regularly but are also required to check the skin for damage and redness at least twice a day, especially the fragile bone area of patients \cite{7}. 
    
At present, there are many different kinds of products on the market that can provide auxiliary support to minimize the risk of patients acquiring pressure injuries \cite{7}. First of all, the most essential products belong to a category of static surface products that do not need to be connected to an additional energy supply, such as foam-filled mats, air mats and liquid mats. This type of mat relieves the pressure by reducing the pressure on the hinge joint \cite{11} of the human body to prevent the direct damage caused by the bed to the patient's skin. The second type of mat can be called dynamic surface mats, such as alternating air mats and pneumatic ripple mats. As the most representative pressure injury prevention product, these pneumatic mats are composed of multiple air chambers that can be inflated and deflated independently, and this function decided that this product should connect with external electricity and air pumps to ensure the continuous operation of those air chambers so that those air chambers can be cyclically and alternately inflated and deflated to redistribute the pressure on the skin surface of the patient \cite{7}. Because of the better pressure injury prevention effect and affordable cost of pneumatic mats, this dynamic surface mat has become one of the most popular products to prevent stress injuries in hospitals and nursing homes worldwide \cite{7}.
    
    \vspace*{1\baselineskip}
    \begin{figure}[h]  

    \centering  

    \includegraphics[width=0.6\linewidth]{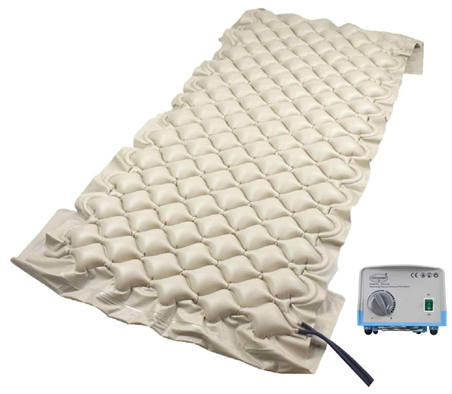} 

    \caption{One type of air-filled mat \cite{7}.}

    \end{figure}    
    
With the development of electronic technology and civilian machinery in recent years, many similar products combine computers and mechanically controlled nursing beds to develop intelligent anti-decubitus beds. The principle is to help patients turn over or move by themselves and nursing staff through mechanical control. However, these products are generally large in size and expensive and often not particularly ideal for the actual effect of pressure injuries prevention. 

\subsection{Smart mat}

The prevention of pressure injury as an inconspicuous process in the medical care process often brings tremendous difficulty to the care process to medical staff. The high occurrence possibility is more likely to cause physical and economic burdens to patients and their families. However, the current products on the market used to prevent pressure injuries have apparent defects. Some mats that use traditional air chambers to relieve the local skin pressure of the patient cannot intuitively and accurately predict the exact position of the patient's oppressed skin. Moreover, mechanical-assisted hospital beds can assist patients to process some movements, like turning over, according to the detecting data, but due to high manufacture budget and large size, those products are rarely applied to every patient who is suffering from pressure injuries \cite{7}. 

Therefore, this work is dedicated to the development of a lightweight smart mat that can directly detect the pressure distribution of the patient’s body. 

\subsubsection{Aim of the product}

(1) This mat should meet the requirements of lightness and be easily portable.	

(2)	The smart mat can accurately detect the pressure distribution of the patient's body through the sensor matrix and visualize it through the PC.

(3)	The alarm function of the smart mat can remind the medical staff to help the patient complete the turning over after detecting that a particular part of the patient's body has been compressed for a long time.

(4)	Abundant health indicators detection such as breathing rate, temperature and humidity, as an additional function to enrich the intelligence of the entire mat.

\section{Pressure-sensitive mat}

Based on existing research, we are looking for a pressure-sensitive bed sheet that can detect the pressure on the patient. The most basic function of the bedsheet is to be able to clearly display the patient's pressure distribution image so that the nursing staff can intuitively observe the patient's body stress distribution to better predict and avoid pressure injuries. Therefore, the production of the smart mat cannot be separated from the algorithm and design of a large number of high-resolution pressure sensor matrices.

\subsection{Structure design of pressure-sensitive devices}

In \cite{13}, the authors proposed a more creative and affordable solution to detect the state and pressure of a person when sitting or lying down. In terms of system architecture, they proposed a cheap and inconspicuous structural model of textile pressure sensing matrix. The pressure sensing area is used to collect information through the polymer foam as a pressure-sensitive element between two electrodes, and the density of the conductive material in the foam will increase when the subject is pressured, which will be resulting in a decrease in resistance, and then the collected resistance signal is digitized by an analog-to-digital converter (ADC). In addition, the sensors arranged in a matrix will create a grayscale image with the collected pressure information, which shows the shape and size of the pressure area and generates a pressure image containing a large number of details of human activities through a high-precision 24-bit ADC sampling rate \cite{16}. The sensors of their pressure-sensitive mat are connected by using uniform parallel n×m metal strips, where n and m are parallel and vertical metal strips respectively, and the contact point of each metal strip can appear as a pixel in the grayscale image \cite{16}.

\begin{figure}[h]  

    \centering  

    \includegraphics[width=0.8\linewidth]{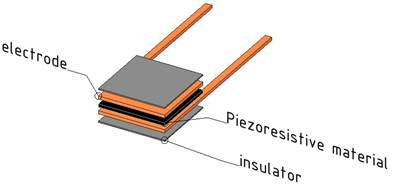} 
    
    \caption{Structure of force-sensor resistor \cite{12}.}
 \label{fig:FSR}
    \end{figure}

The reliability of data obtained by a sensor matrix mainly depends on the performance of individual pixels and the overall pixel resolution. First of all, the structure of a single sensor that is usually designed is similar to the sandwich shape depicted in fig. \ref{fig:FSR}.

    \vspace*{1\baselineskip}
    
According to the structure in Fig. \ref{fig:FSR}, the sensor on a single pixel is composed of piezoelectric materials, electrodes and spacers. The operation responsibility of piezoelectric material require it to feel the pressure change and convert the data into a detectable resistance value, the electrode should supply the voltage to the piezoelectric material and gather the detected resistance change to data processing unit, finally clamped on the two outermost insulators of the whole structure to reduce the interference caused by voltage changes to confirm the transmitted data accurately \cite{12}.

On the other hand, Jose et al. also proposed a point pressure mapping system based on a single-layer textile sensor \cite{17}. In this design, the silver-plated yarns are sewn on the cotton fabric in parallel, and the yarns and cotton fabrics are sewn vertically and horizontally to make the yarns and the cotton fabric present a net shape, and then the conductive polymer solution is applied to the junction of each yarn. When pressure is felt at the processed wiring, a resistance change is generated and the information is sent to a display device \cite{17}. 
There are a variety of smart mat systems used in various scenarios, which collect pressure data using matrix-arranged varistors and then perform analog-to-digital conversion via cables linking each resistor.


No matter how the materials of electrode and sensor units are selected, the working principle of the entire resistance matrix is roughly the same. All the individual piezoelectric sensors connected in series with the electrodes constitute the entire pressure detection system, and the number of piezoelectric sensors per unit area determines the resolution of the system. 




    
    
\subsection{Data acquisition and algorithm}

In \cite{13}, the authors used the Field programmable gate array (FPGA) based processor to control the switch array and the analog-to-digital converter to collect the sample data in the resistance matrix by scanning each row \cite{13}. FPGA is an integrated circuit that contains a large number of logic cells. It supports multiple I/O protocols and logic cell arrays through programming to achieve complex functions, such as digital signal processing, communication processing, image and video processing, etc. The FPGA controls the switches in each column of the matrix to be turned on in turn, and then the voltage is passed to the multiplexer for further analog-to-digital conversion. The scanned image is connected to generate an $n\times m$ pressure image. 

    \vspace*{1\baselineskip}
    \begin{figure}[h]  

    \centering  

    \includegraphics[width=0.9\linewidth]{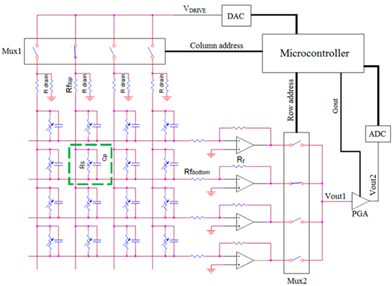} 
    
    \caption{Force-sensor matrix circuit \cite{15}.}
\label{Fig:FSMC}
    \end{figure}  
    
In order to obtain the pressure signal, Jose ‘s team \cite{15} also used similar multiplexing electronic equipment. In each row and column of the resistance matrix, a microcontroller and an amplifier were used to collect information, which is amplified and uploaded to Analog to digital conversion (ADC) through the multiplexer and then  processed by microcontroller. At the same time, in order to reduce the crosstalk between the resistors, it is necessary to ground each row and column resistors \cite{15}. The diagram is shown in Fig, \ref{Fig:FSMC}.

Many existing studies have shown the relationship between the resistance of the piezoresistive material and the force between the two electrodes \cite{12}, and the resistor includes the piezoresistive material will be called as force-sensor resistor (FSR), the following formula can express the relationship between the resistance and pressure force.

\[ R_{FSR} =\frac{ {\rho*K} }{F} \]

\begin{figure}[h]  

    \centering  

    \includegraphics[width=0.4\linewidth]{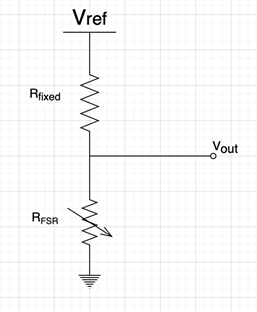} 
    
    \caption{working principle of FSR based on Ohm's law}
\label{Fig:Ohm}
    \end{figure}

In this formula, the resistance value of the FSR is inversely proportional to the force received. When the resistivity of contacting surface R and the roughness elastic properties of surface K is constant for one material, the resistance would decrease if the force exerted on the piezoresistive material keeps increasing.  Therefore, according to the principle of voltage division and variable resistance of piezoresistive materials, the basis of the circuit can be constructed through the image shown in Fig. \ref{Fig:Ohm}.

    \vspace*{1\baselineskip}

The sensor \(R_{FSR}\) is simply connected in series with the fixed resistor \(R_{fixed}\), and the voltage measured at \(V_{out}\) can be obtained by Ohm's law, so the specific calculation is shown below.

\[ V_{out} = V_{in}*( \frac{ R_{FSR} }{ R_{fixed}+ R_{FSR}}) \]

If the equation of R and pressure will be substituted into the above formula, the relationship between the output voltage and pressure force can be expressed as follows.

\[ V_{out} = V_{in}*( \frac{ \rho*K } { F*R_{fixed}+ \rho*K } )\]

Therefore, the pressure is inversely proportional to the output voltage, which means that the value of the output voltage will be smaller when the pressure continues to increase.

\begin{figure}[h]  

    \centering  

    \includegraphics[width=0.6\linewidth]{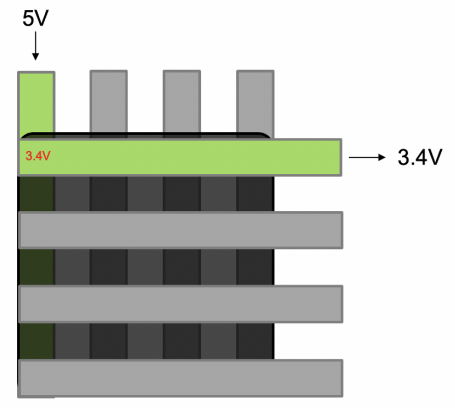} 
    
    \caption{Single piece FSR working in the matrix}
\label{Fig:FSR_voltage1}
    \end{figure}

\begin{figure}[h]  

    \centering  

    \includegraphics[width=0.6\linewidth]{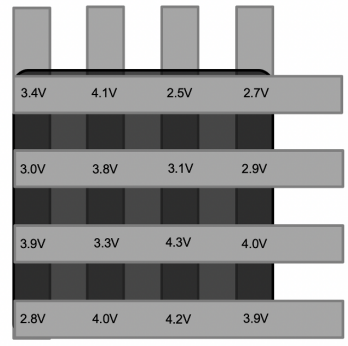} 
    
    \caption{Single piece FSR working in the matrix}
\label{Fig:FSR_voltage2}
    \end{figure}

After combining some previous cases and experiences, the primary working status of the pressure detection mat based on the FSR resistance matrix and the voltage distribution of each resistance during operation is shown in Figs. \ref{Fig:FSR_voltage1} and \ref{Fig:FSR_voltage2}.
    \vspace*{1\baselineskip}
    
    \vspace*{1\baselineskip}
    
Figs. \ref{Fig:FSR_voltage1} and \ref{Fig:FSR_voltage2} clearly show that when the voltage of 5V output by the row electrode is added to a single FSR, the output voltage value changes due to the voltage division principle as the data of a single pixel. Suppose that the output of each pixel on the entire mat is arranged successively according to the original position. In that case, the data table generated over the whole mat due to uneven pressure distribution is shown in Fig. \ref{Fig:FSR_voltage2}. However, when the resolution of the mat is substantial, a large amount of data will cause a significant increase in information processing load, so the multiplexer will be used to control the voltage selective input and output.

\begin{figure}[h]  

    \centering  

    \includegraphics[width=1\linewidth]{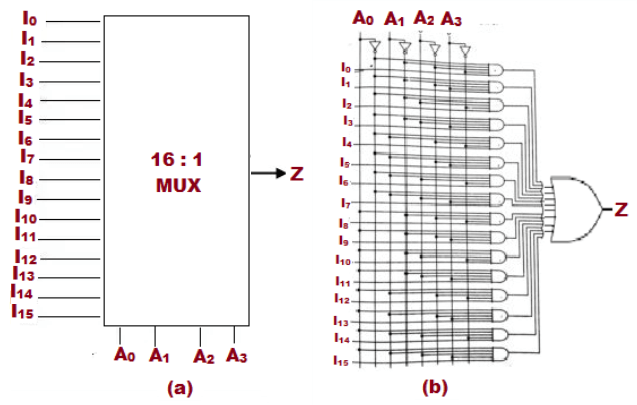} 
    
    \caption{16 channel multiplexer}
\label{Fig:multiplexer}
    \end{figure}

The multiplexer shown in Fig. \ref{Fig:multiplexer} can control the analog signal output in 16 channels through 4 digital ports, thus it will significantly reduce the number of ports used in Arduino so that to simplify the circuit.The image's composition is realized by the combination of multiple pixels, so for a picture, the increasing number of pixels can bring the high quality of the picture. As shown in Fig. \ref{Fig:FSR_voltage2}, the voltage value of each FSR sensor caused by changing resistance could be regarded as every pixel of the picture, and a completed image can be produced by integrating the real-time data of all sensors of the entire mat. Therefore, although the abundant number of sensors are more likely to plot the accurate image of the pressure distribution, it also increases the number of electrodes connected to each sensor, thereby increasing the complexity of the circuit and the difficulty of data processing. Therefore, a multiplexer that can provide multiple input and output paths can simplify the circuit to transmit data more conveniently. 

    \vspace*{1\baselineskip}



\section{Design and Methodology \label{cha:methods}}

Through the investigation and understanding in the previous section, the suitable materials and structures according to the requirements of the products should be selected accurately. Therefore, this section mainly introduces the analysis of different materials and structural designs.

\subsection{Materials}


\subsubsection{Pressure sensor}

\begin{figure}[h]  

    \centering  

    \includegraphics[width=1\linewidth]{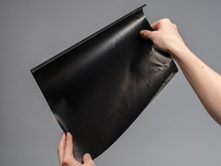} 
    
    \caption{Velostat}
\label{Fig:velostat}
    \end{figure}

Basically, the most important part of a smart mat is the ability to detect pressure and output the signals. Like most previous work, the easiest way to feel pressure is to use a pressure-sensitive sensor, but due to the invisibility required by smart sheets, due to the complexity of the structure, we did not use the polymer foam piezoelectric sensor used in \cite{13}. Instead, we use a pressure-sensitive conductive sheet called Velostat as shown in Fig. \ref{Fig:velostat}, which is a conductive polymer foil made of polyolefin, and it is often used to protect equipment and objects that are easily damaged by classic electrical discharges.

    \vspace*{1\baselineskip}
    
The pressure-sensitive properties of this material reduce the resistance in the squeezed and pressured state, and the excellent softness and flexibility of this material keep it from being damaged under heavy pressure.

In addition, we have done some calculations to verify that the pressure cushion needs to detect the approximate pressure range of various parts of the human body on the bed. Assuming an adult’s weight 80kg and the contact area between the body and the bed is about 170*30cm. 
Average pressure of people lying flat: \[ 80kg/{170*30}cm^2=0.015kg/cm^2=0.22psi\]
Pressure of a person standing: \[80kg/350cm^2=0.22kg/cm^2=3.13psi\]

\begin{figure}[h]  

    \centering  

    \includegraphics[width=1\linewidth]{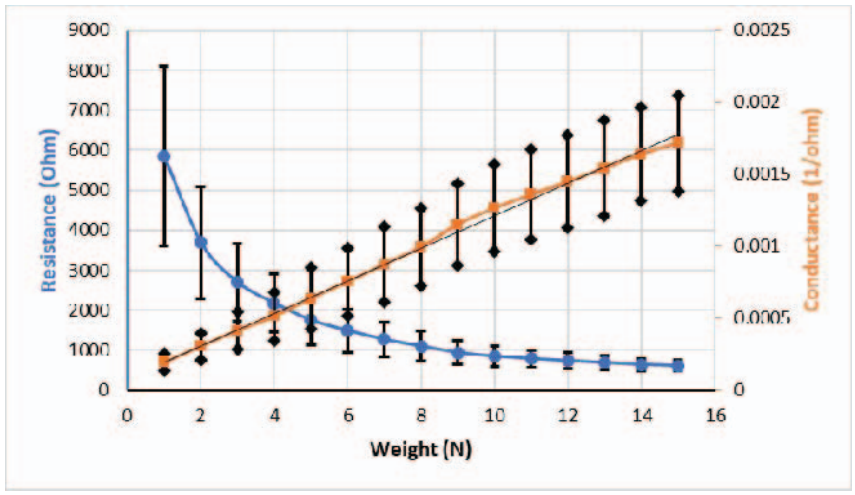} 
    
    \caption{Velostat resistance vs. pressure curve \cite{23}.}
\label{Fig:velostat_resistance}
    \end{figure}

The pressure sensor selection needs to consider the range of the sensor's resistance changing with pressure. From the test experiment conducted by Valle-Lopera on the Velostat material \cite{12}, it can be known that the resistance change curve of the sensor measured when a force ranging from 0 to 500N is applied to the velostat is shown in Fig. \ref{Fig:velostat_resistance}.

    \vspace*{1\baselineskip}
    
The pressure range that causes the change of the resistance of the velostat material includes the pressure generated by the human body to the ground or the bed. Thus this material is one of the appropriate materials that can be applied. 

\subsubsection{The base material of mat}

Since the bed sheet may be in direct contact with the patient's skin, the material selected must be breathable and harmless, and it is desirable to provide an insulated environment to ensure the normal operation of the sensor. Hence, the fabric we chose is a ripstop nylon fabric, which is breathable, non-toxic and waterproof, and is very suitable as a basic material for smart pads.

\subsubsection{Other materials selection }

The development board selected for this experiment is Duinotech MEGA 2560 r3 Arduino, and the flash memory of the development board is 256KB, which is eight times larger than that of the traditional Arduino development board. Furthermore, the conductive material used to connect the sensor matrix is a conductive nylon fabric tape. Because of the nylon material, this tape will not break due to bending and folding, moreover, the electrode used to connect the sensor matrix could be the selected conductive nylon fabric tape, and because of the nylon material, this tape will not break when the mat is bending and folding. The characteristic of high conductivity allows only a few ohms per inch of resistance, and the most convenient thing is that it can be directly attached to one side of the fabric \cite{17}. In addition, the ribbon cable can upload data collected by each electrode to the multiplexer, and the multiplexer controls data collection and sends it to the Arduino for data processing.

\subsection{Structure}

The main structure of the entire mat can be divided into two parts: the mat structure and the circuit structure. The main task of the mat part is to collect the pressure information acting on each FSR and then upload it to the chip. The primary function of the circuit part, including the multiplexer and the chip, is to process the collected data and further transmit it to the PC to prepare for data visualization.

\subsubsection{Structure of sensor area}

\begin{figure}[h]  

    \centering  

    \includegraphics[width=0.5\linewidth]{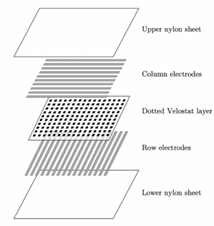} 
    
    \caption{Three main layers of mat.}
\label{Fig:3layer}
    \end{figure}

One type of three-layer structure used in the sensor matrix mat mentioned in many previous works \cite{12,15,16} is adopted in this mat. It can be clearly seen from Fig. \ref{Fig:3layer} that the row electrode layer and the column electrode layer is located on top and bottom respectively in the entire mat, and the middle layer (dotted velostat layer) is the sensing area. The cloth selected in the above-mentioned materials can be used as the wrapping layer of the entire mat because it requires to prevent interference from the external environment and contact the human body directly. Furthermore, each wire of the two ribbon cables is connected to every electrode in the row and column to upload the information received by the electrodes.
    \vspace*{1\baselineskip}
\subsubsection{Circuit diagram}

The circuit connection applied to the smart mat for receiving is mainly based on the resistance divider of Ohm's theorem (see Appendix A). A circuit diagram containing all the multiplexer and Arduino pin ports presents the circuit for a mat of $128\times 68$ resolution, and all connections and explanations have been included in the diagram.

\subsection{Information visualization}

\subsubsection{Cross talk removal}

The occurrence of crosstalk problems is more likely to affect the normal operation of the system. For example, medical staff may receive false alarm information when the pressure display is greater than the actual value due to crosstalk. There are many possible causes of crosstalk, for example, when the FSR is arranged in a matrix form, it will have the side effect of crosstalk because there is no electrical isolation between each sensor. In this case, the intersection of the row and the column will cause some current to be conducted from one sensor to the other through the electrodes \cite{18}.

    \vspace*{1\baselineskip}
    \begin{figure}[htb]  

    \centering  

    \includegraphics[width=0.5\linewidth]{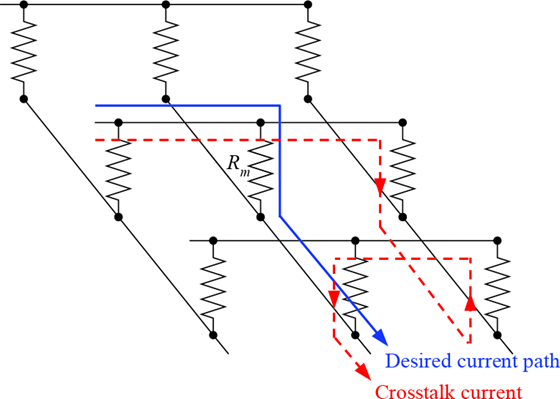} 
    
    \caption{Crosstalk path in sensor matrix \cite{18}.}
\label{Fig:crosstalk}
    \end{figure} 
It can be seen from Fig. \ref{Fig:crosstalk} that the start and endpoints of each resistor are connected to others, which means that there is another choice of the path through the current instead of flowing the current according to the initially expected path. Under the intense pressure on a large number of sensors, these currents passed by each electrode will flow through these selectable paths instead of being conducted from the designated line alone, which will cause poor image quality. Please note that crosstalk often does not appear in a small range and low-pressure condition \cite{18}.

Crosstalk may be eliminated by software and/or hardware. In terms of hardware, the zero potential circuit configuration using the operational amplifier and diode could solve the crosstalk problem. On the other hand, the predicted value of Convolutional Neural Network (CNN) in software may also minimize the crosstalk to a certain extent \cite{27}. In addition, deep learning algorithms and denoising techniques developed in our existing works, e.g., \cite{IREALCARE1,IREALCARE2,IREALCARE3,IREALCARE4,IREALCARE5} can be used to mitigate the crosstalk issues. No matter which method is adopted, there are pros and cons in different aspects.

The principle of using diodes is to add diodes in series with each sensor to block current from flowing through unspecified paths \cite{27}. According to this method, the crosstalk problem can be effectively solved without increasing the cost of the smart mat. However, there are currently no flexible diodes on the market, which means that every metal diode will force the patient's body. First, the patient will have a strong sense of foreign body, and second, the patient will feel strong pressure at the position where the diode appears, which may further lead to the appearance of pressure sores \cite{26}.



    


The operational amplifier can solve the pressure problem caused by the diode mentioned above, because it can be placed outside the matrix together with other electronic devices, which will not bring uncomfortable experience to the patient \cite{15}. Because the operational amplifier has zero potential, it can prevent the current from passing through the sensor that is not under pressure and prevent the row or column that is not being measured from changing \cite{15}. The specific implementation method is to use an operational amplifier with feedback in the transimpedance configuration of each row output. The principle is to use the principle of virtual grounding to ground one of the inputs of the operational amplifier and input a voltage of 0 volts at the other end to construct a virtual ground. However, this solution requires a large number of operational amplifiers, although it will not affect the daily life of the patient in the hospital bed \cite{27}.

In the software, intelligent algorithms and new software systems can be used to eliminate crosstalk. Among them, the convolutional neural network can predict the shape of the object by using the original output value with crosstalk \cite{26}. However, in order to ensure the reliability and authenticity of the results, a large amount of raw data needs to be collected and compared with the control group, which may take a long time and increase the cost of the mat \cite{26}.

Another algorithm in software to reduce crosstalk is the median filter. In order to minimise the crosstalk between the various resistances between the grids, first, the median filter algorithm needs to be applied to the hardware processing, and in order to ensure the robustness of the data processing, the median filter can also be used directly to filter the data. Second, the output of each pixel point contains the DC current brought by the data collection process, so the removal of the DC current can reduce the value occupied in the matrix in the outgoing circuit. Finally, bilinear interpolation to sample the intermediate pixels can improve the overall pixel resolution \cite{27}. Among them, the median filter, as a kind of non-linear filter, can effectively remove the impulse noise in the image, and the principle is to replace the grayscale of each pixel with the median grayscale in the pixel neighbourhood \cite{15}. In addition, the bilinear interpolation method used in sampling is to calculate each missing pixel through the weighted average of the pixels on the adjacent boundary so that the entire image becomes coherent and smooth \cite{15}.

\subsubsection{Posture detection }

When applied to the actual scene, a smart mat can detect the shape of the human body to determine the position and posture. As one of the most commonly used interactive interfaces for patients, embedded sensors can realize the mat to extract rich sensory information \cite{16}.

At present, the function of most sensors is based on time-domain data analysis of acquired sensing signals, usually by signal amplitude and frequency. The smart mat can recognize various patient postures through the array, such as lying on the side, lying on the back or prone posture. According to the overall amplitude of the output signal, the force points and position changes of different parts of the body can be easily identified according to the colour. For example, the colour of the most stressed areas is usually red, and the parts of the body that are less stressed are often green or blue \cite{16}. 

\subsubsection{Respiratory rate}

Through filtering and spectrum analysis methods, the patient's breathing frequency can be roughly predicted. Each induction point can collect the breathing signal from the patient's pressure changes through the conductance summation. Through filtering and spectrum analysis methods, the patient's breathing frequency can be roughly predicted. Each induction point can collect the breathing signal from the patient's pressure change through conductance summation and use a filter to process the signal and perform Fourier transform to extract the signal more clearly and identify the frequency with the highest amplitude. Among them, the highest frequency is the calculated respiratory frequency \cite{28}. Please note that other methods can also be used for posture and respiratory rate detection, such as, WIFI based methods \cite{IREALCARE5}, and radar imaging based techniques, \cite{GI1,GI2,GI3,GI4}.


\section{Results and Analysis\label{sec:V}}

\subsection{Tests and Results\label{cha:results}}

Before starting to work on the prototype, we have done some tests to analyse factors affecting performance of pressure sensors. The test on the surface of the conductive tape is divided into three steps. The first step is to test the pressure value of the unpressured resistance and the pressured resistance of the two adhesive sides. Under this test, data measured in this set of experiments are all 4k ohms. The second is to test an adhesive side and a normal side. The values of the unpressured resistance and the pressured resistance are 20 and 1.6 respectively. The last step is a pressure test on two normal surfaces, where the unpressured resistance is still 20, but the pressured resistance is significantly reduced to 0.02. Through this series of tests, it can be seen that the effect of two normal sides changes is the most obvious one.

\begin{figure}[htb]  

    \centering  

    \includegraphics[width=0.6\linewidth]{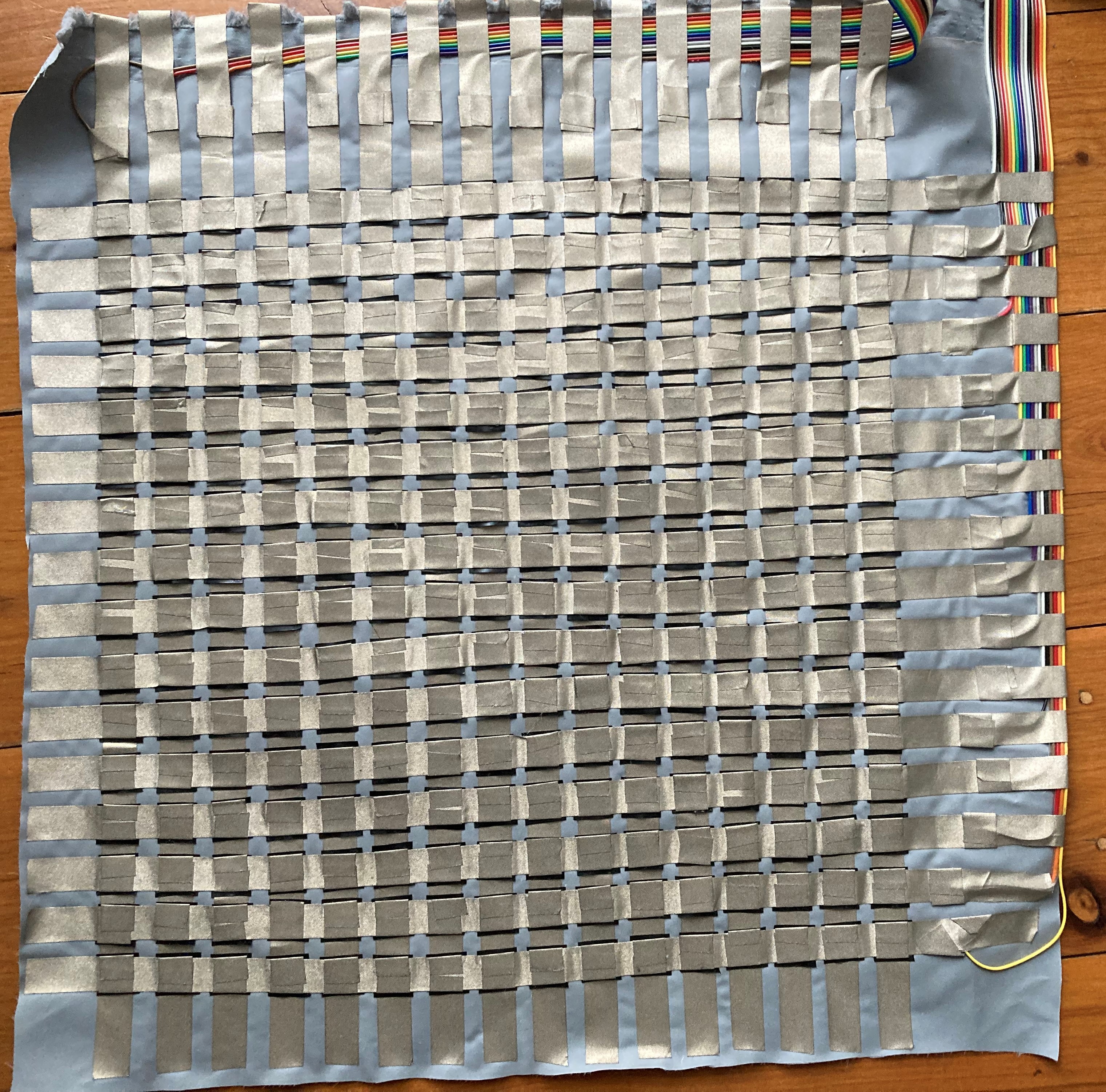} 
    
    \caption{prototype}
\label{Fig:prototype}
    \end{figure}

It can be seen from the test result that the resistance variable characteristics of Velostat are completely invalid when two adhesive surfaces are used to connect with Velostat directly. Therefore, for making the prototype, we avoid direct contact between the adhesive surface of the electrode and the velostat.  
The prototype of the mat is shown in Fig. \ref{Fig:prototype}.


    \vspace*{1\baselineskip}

\begin{figure}[htb]  

    \centering  

    \includegraphics[width=1\linewidth]{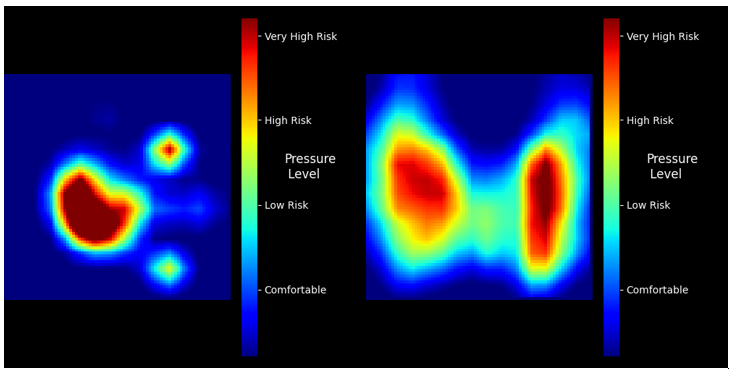} 
    
    \caption{Image display of hands and buttocks}
\label{Fig:hands}
    \end{figure}

\begin{figure}[htb]  

    \centering  

    \includegraphics[width=1\linewidth]{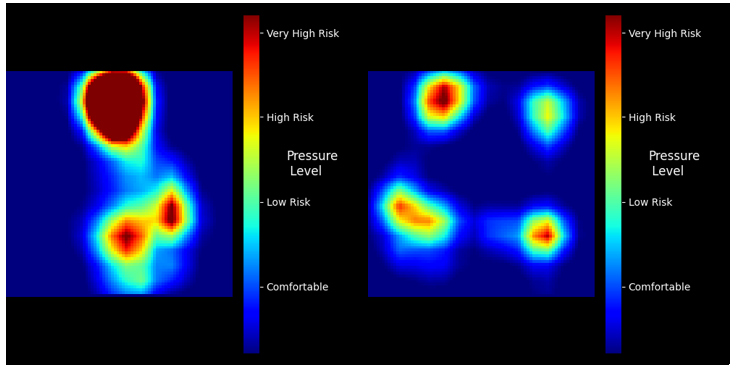} 
    
    \caption{Image display of one foot and both feet}
\label{Fig:feet}
    \end{figure}

\begin{figure}[htb]  

    \centering  

    \includegraphics[width=1\linewidth]{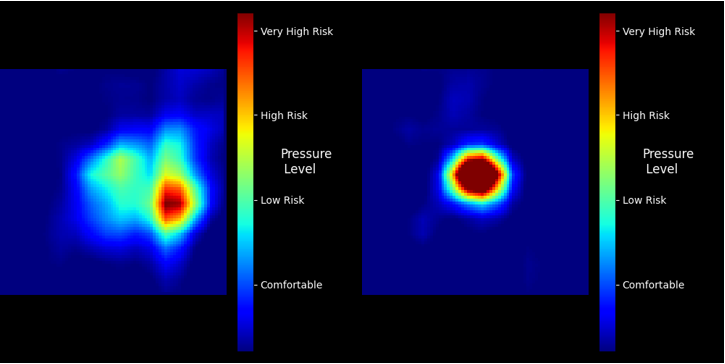} 
    
    \caption{Image display of face and chin}
\label{Fig:face}
    \end{figure}

With the prototype of the mat, we tested the pressure distributions for different parts of human body. The results for hands, feet and face are shown in Figs. \ref{Fig:hands}, \ref{Fig:feet}, and \ref{Fig:face}, respectively.
    \vspace*{1\baselineskip}
    
    \vspace*{1\baselineskip}
    
    \vspace*{1\baselineskip}

From these figures, we can see that the image display is very clear, and the image also has an alarm function after long-term pressure. The pressure on the area is divided by the use of color bars, and the image can clearly show which area is under more pressure. Moreover, At this time, the measured resistance value of the single resistance in the red area is below 1000 ohms. From Fig. \ref{Fig:velostat_resistance}, we can see that the pressure per unit area in the red area is more than ten kilograms. 

\section{Summary}



In this work, we developed a smart mat for monitoring pressure injury development. We built a sensor matrix over the mat using Velostat as a force sensor resistance (FSR) to collect the patient's body's pressure distribution. We then uploaded the processed distribution information to the PC for data display using Arduino. The goal of this effort is to reduce the pressure injuries that medical staff suffers while treating and caring for patients during the pandemic. In the current  design, the size of the mat is relatively small, in the future, we will build a large size mat for full body pressure distribution monitoring. Furthermore, privacy concerns for medical data storage \cite{privacy1,privacy2,privacy3,privacy4} 
 and connectivity via different networks, such as cellular networks \cite{NC, UAV_THz, cellular1,cellular2,cellular3,MIMO_capacity,UAVdownlink,OC-isit,codedcpm1,codedcpm2,codedcpm3} and sensor networks \cite{network_capacity,DistributedRateless,DistributedRateless2,DistributedRateless3,NC4,NC5,WRN2,WRN3, Raptor_ML}  
are also our future plan.


\end{document}